\documentclass{nature}

\usepackage{amssymb}
\usepackage{amsmath}
\usepackage{graphicx}

\newcommand{\om}{\omega}
\newcommand{\Om}{\Omega}
\newcommand{\p}{\partial}
\newcommand{\be}{\begin{equation}}
\newcommand{\ee}{\end{equation}}
\newcommand{\bsub}{\begin{subequations}}
\newcommand{\esub}{\end{subequations}}
\newcommand{\bea}{\begin{eqnarray}}
\newcommand{\eea}{\end{eqnarray}}
\newcommand{\bi} {\begin{itemize}}
\newcommand{\ei} {\end{itemize}}
\newcommand{\bmat} {\begin{pmatrix}}
\newcommand{\emat} {\end{pmatrix}} 

\renewcommand{\vec}[1]{\mathbf{#1}}

\title{Observation of superradiance in a vortex flow}

\author{Theo Torres$^{1}$, Sam Patrick$^{1}$, Antonin Coutant$^{1}$, Mauricio Richartz$^{2}$, Edmund W.~Tedford$^{3}$ \&  Silke Weinfurtner$^{1,4}$}

\begin{document}

\maketitle

\begin{affiliations}
 \item School of Mathematical Sciences, University of Nottingham, University Park, Nottingham, NG7 2RD, UK
 \item Centro de Matem\'atica, Computa\c c\~ao e Cogni\c c\~ao,
Universidade Federal do ABC (UFABC), 09210-170 Santo Andr\'e, S\~ao Paulo, Brazil
\item Department of Civil Engineering, University of British Columbia, 6250 Applied Science Lane, Vancouver, Canada V6T 1Z4
\item	School of Physics and Astronomy, University of Nottingham, Nottingham, NG7 2RD, UK.
\end{affiliations}


\begin{abstract}
Wave scattering phenomena are ubiquitous to almost all Sciences, from Biology to Physics. When an incident wave scatters off of an obstacle, it is partially reflected and partially transmitted. Since the scatterer absorbs part of the incident energy, the reflected wave carries less energy than the incident one. However, if the obstacle is rotating, this process can be reversed and waves can be amplified, extracting energy from the scatterer. Even though this phenomenon, known as superradiance~\cite{Brito:2015oca,Richartz:2009mi}, has been thoroughly analysed in several theoretical scenarios (from eletromagnetic radiation scattering on a rotating cylinder~\cite{zeldovich1,zeldovich2} to gravitational waves incident upon a rotating black hole~\cite{misner,staro1,staro2}), it has never been observed. Here we describe in detail the first laboratory detection of superradiance. We observed that plane waves propagating on the surface of water are amplified after being scattered by a draining vortex. The maximum amplification measured in the experiment was 20\%, obtained for 3.70~{$\mathbf{Hz}$} waves, in a 6.25~{$\mathbf{cm}$} deep fluid.  
Our results are consistent with superradiant scattering caused by rapid rotation. 
In particular, a draining fluid can transfer part of its rotational energy to incident low-frequency waves. Our experimental findings will shed new light on Black Hole Physics, since shallow water waves scattering on a draining fluid constitute an analogue of a black hole~\cite{unruh,weinfurtner,steinhauer1,steinhauer2,dolan3}. We believe, especially in view of the recent observations of gravitational waves~\cite{gwaves}, that our results will motivate further research (both theoretical and experimental) on the observation of superradiance of gravitational waves.

\end{abstract}

In water, perturbations of the free surface manifest themselves by a small change $\xi(t,\vec x)$ of the water height. On a flat bottom, and in the absence of flow, linear perturbations are well described by superpositions of plane waves of definite frequency $f$ and wave-vector $\vec k$. When surface waves propagate on a changing flow, the surface elevation is generally described by the sum of two contributions $\xi = \xi_I + \xi_S$, where $\xi_I$ is the \emph{incident} wave produced by a source, e.g.~a wave generator, while $\xi_S$ is the \emph{scattered} wave, generated by the interaction between the incident wave and the background flow. In this work, we are interested on the properties of this scattering on a draining vortex flow which is assumed to be \emph{axisymmetric} and \emph{stationary}. At the free surface, the velocity field is given in cylindrical coordinates by $\vec v = v_r \vec e_r + v_\theta \vec e_\theta + v_z \vec e_z$.

Due to the symmetry, it is appropriate to describe $\xi_I$ and $\xi_S$ using polar coordinates $(r, \theta)$. Any wave $\xi(t,r,\theta)$ can be decomposed into partial waves~\cite{dolan3,Newton}, 
\be \label{Fourier_t_th}
 \xi(t,r,\theta) = \sum_{m \in \mathbb Z} \int \varphi_{f, m}(r) \frac{e^{-2 i \pi f t + i m \theta}}{\sqrt{r}} df, 
\ee
where $m \in \mathbb Z$ is the azimuthal number and $\varphi_{f, m}(r)$ denotes the radial part of the wave. Each component of this decomposition has a fixed angular momentum proportional to $m$, instead of a fixed wave-vector $\vec k$. (To simplify notation, we drop the indices ${}_{f, m}$ in the following.) Since the background is stationary and axisymmetric, waves with different $f$ and $m$ propagate independently. Far from the centre of the vortex, the flow is very slow, 
and the radial part $\varphi(r)$ becomes a sum of oscillatory solutions,  
\be \label{in_out}
\varphi(r) = A_{\rm in} e^{-i k r} + A_{\rm out} e^{i k r},
\ee
where $k = ||\vec k||^2$ is the wave-vector norm. This describes the superposition of an inward wave of (complex) amplitude $A_{\rm in}$ propagating towards the vortex, and an outward one propagating away from it with amplitude $A_{\rm out}$. These coefficients are not independent. The $A_{\rm in}$'s, one for each $f$ and $m$ component, are fixed by the incident part $\xi_I$. If the incident wave is a plane wave $\xi = \xi_0 e^{-2i\pi f t + i \vec k \cdot \vec x}$, then the partial amplitudes are given by $A_{\rm in} = \xi_0 e^{i m \pi + i \pi/4}/\sqrt{2\pi k}$. In other words, a plane wave is a superposition containing \emph{all} azimuthal waves, 
something that we have exploited in our experiment. On the contrary, $A_{\rm out}$ depends on the scattered part $\xi_S$, and how precisely the waves propagate in the centre and interact with the background vortex flow. In the limit of small amplitudes, there is a linear relation between the $A_{\rm in}$'s and $A_{\rm out}$'s, and by the symmetries of the flow, different $f$ and $m$ decouple~\cite{dolan3,Richartz:2014lda}. 

This allows us to define the \emph{reflection coefficient} at fixed $f$ and $m$ as the ratio between the outward $\left(J_{\rm out}\right)$ and inward $\left(J_{\rm in}\right)$ energy fluxes, 
\be
R = \sqrt{\frac{J_{\rm out}}{J_{\rm in}}}. 
\ee
In the linear approximation, the wave energy is a quadratic quantity in wave amplitude, and $R$ is proportional to the amplitude ratio $|A_{\rm out}/A_{\rm in}|$. 

If $|R| < 1$, the wave has lost energy during the scattering, and hence has undergone absorption. In this work we show experimentally that, under certain conditions, the reflection coefficient satisfies $|R| > 1$. We further argue that the amplified wave has extracted rotational energy from the vortex during the process.


We conducted our experiment in a $3~\mathrm{m}$ long and $1.5~\mathrm{m}$ wide rectangular water tank. Water is pumped continuously in from one end corner, and is drained through a hole ($4~\mathrm{cm}$ in diameter) in the middle. The water flows in a closed circuit. 
We first establish a stationary rotating draining flow by setting the flow rate of the pump to $37.5~\pm~0.5~\mathrm{\ell/min}$ and waiting until the depth (away from the vortex) is steady at $6.25 ~\pm ~0.05 ~\mathrm{cm}$. We then generate plane waves from one side of the tank, with an excitation frequency varying from $2.87~\mathrm{Hz}$ to $4.11~\mathrm{Hz}$. On the side of the tank opposite the wave generator, we have placed an absorption beach (we have verified that the amount of reflection from the beach is below $5\%$ in all experiments). We record the free surface with a high speed 3D air-fluid interface sensor. The sensor is a joint-invention\cite{enshape} (patent No.~DE 10 2015 001 365 A1) between The University of Nottingham and EnShape GmbH (Jena, Germany). 

Using this data, we apply two filters. We first perform a Fourier transform in time, in order to single out the signal at the excitation frequency $f_0$. This allows us to filter out the (stationary) background height, lying at $f=0$, as well as the high frequency noise. Moreover, we observe that the second harmonic, at $2 f_0$, is also  excited by the wave generator. This gives us an upper bound on the amount of nonlinearity of the system. In all experiments, the relative amplitude of the second harmonic compared to the fundamental stays below $14 \%$. The obtained pattern shows a stationary wave of frequency $f_0$ scattering on the vortex, which consists of the interfering superposition of the incident wave $\xi_I$ with the scattered one $\xi_S$. This pattern is shown on \textbf{Fig.~1} 
for various frequencies, and looks very close to what was observed numerically in simple bathtub flow models~\cite{dolan3}. We also observe that incident waves have more wave fronts on the upper half of the vortex in comparison with the lower half, see various wave characteristics in panels (A-F) in \textbf{Fig.~1}. 
This angular phase shift is analogous to the Aharonov-Bohm effect, and has been observed in previous water wave experiments~\cite{Berry,Vivanco99}. Our detection method allows for a very clear visualization of this effect. 

The second filter is the polar Fourier transform, which selects a specific azimuthal number $m$, and allows the radial profile $\varphi(r)$ to be determined. To extract the reflection coefficient, we use a windowed Fourier transform of the radial profile $\varphi(r)$. The windowing is done on the interval $[r_{\rm min}, r_{\rm max}]$. When $r_{\rm min}$ is large enough, the radial profile $\varphi$ contains two Fourier components [see Eq.~\eqref{in_out}], one of negative $k$ (inward wave), and one of positive $k$ (outward wave). The ratio between their two amplitudes gives us the reflection coefficient (up to the energy correction, see Methods - Wave energy). In order to better resolve the two peaks, we have applied a Hamming window on the radial profile over the interval $[r_{\rm min}, r_{\rm max}]$. In all experiments, $r_{\rm min} \simeq 0.15~\mathrm{m}$, while $r_{\rm max} \simeq 0.39~\mathrm{m}$. We also point out that the minimum radius such that the radial profile reduces to Eq.~\eqref{in_out} increases with $m$. With the size of our window, and the wavelength range of the experiment, we can resolve with confidence  $m=-2,-1,0,1,2$. 

On \textbf{Fig.~2} 
we represent, for several azimuthal numbers $m$, the absolute value of the reflection coefficient $R$ as a function of the frequency $f$. We observe two distinct behaviours, depending on the sign of $m$. Negative $m$'s (waves counterrotating with the vortex) have a low reflection coefficient, which means that they are essentially absorbed in the vortex hole. On the other hand, positive $m$'s have a reflection coefficient close to 1. In some cases this reflection is above one, meaning that the corresponding mode has been amplified while scattering on the vortex. To confirm this amplification we have repeated the same experiment 15 times at the frequency $f = 3.8~\mathrm{Hz}$ and water height $h_0 = 6.25 ~\pm ~0.05 ~\mathrm{cm}$, for which the amplification was the highest. We present the result on \textbf{Fig.~3}. 
On this figure we clearly observe that the modes $m=1$ and $m = 2$ are amplified by factors $R_{m=1} \sim 1.10 \pm 0.06$, and $R_{m=2} \sim 1.15 \pm 0.09$ respectively. On \textbf{Figs. 2}  
and \textbf{3}, 
we have also shown the reflection coefficients obtained for a plane wave propagating on standing water of the same depth. Unlike what happens in the presence of a vortex, the reflection coefficients are all below 1 (within error bars). For low frequencies it is close to 1, meaning that the wave is propagating without losses, while for higher frequencies it decreases due to a loss of energy during the propagation, i.e.~damping. 

The origin of this amplification can be explained by the presence of negative energy waves \cite{Stepanyants, Coutant:2012mf}. Negative energy waves are excitations that lower the energy of the whole system (i.e.~background flow and excitation) instead of increasing it. In our case, the sign of the energy of a wave is  given by the angular frequency in the fluid frame $\om_{\rm fluid}$. If the fluid rotates with an angular velocity $\Om(r)$, we have $\om_{\rm fluid} = 2\pi f - m\Om(r)$. At fixed frequency, when the fluid rotates fast enough, the energy becomes negative. If part of the wave is absorbed in the hole, carrying negative energy, the reflected part must come out with a higher positive energy to ensure conservation of the total energy~\cite{Richartz:2009mi}. Using Particle Imaging Velocimetry (PIV), we have measured the velocity field of the vortex flow of our experiment. As we see on \textbf{Fig.~4A}, 
close to the centre, the angular velocity is quite high, and the superradiant condition $2 \pi f < m \Omega$ is therefore satisfied for our frequency range. 


Our experiment demonstrates that a wave scattering on a rotating vortex flow can carry away more energy than the incident wave brings in. Our results show that the phenomenon of superradiance is very robust and requires few ingredients to occur, namely high angular velocities, allowing for negative energy waves, and a mechanism to absorb these negative energies. For about half of the frequency range, our results confirm superradiant amplification despite a significant damping of the waves. The present experiment does not reveal the mechanism behind the absorption of the negative energies. The likely possibilities are that they are dissipated away in the vortex throat, in analogy to superradiant cylinders~\cite{zeldovich2,Cardoso:2016zvz}, that they are trapped in the hole~\cite{Richartz:2009mi} and unable to escape, similarly to what happens in black holes~\cite{misner}, or a combination of both. A possible way to distinguish between the two in future experiments would be to measure the amount of energy going down the throat.

\begin{center}
\includegraphics{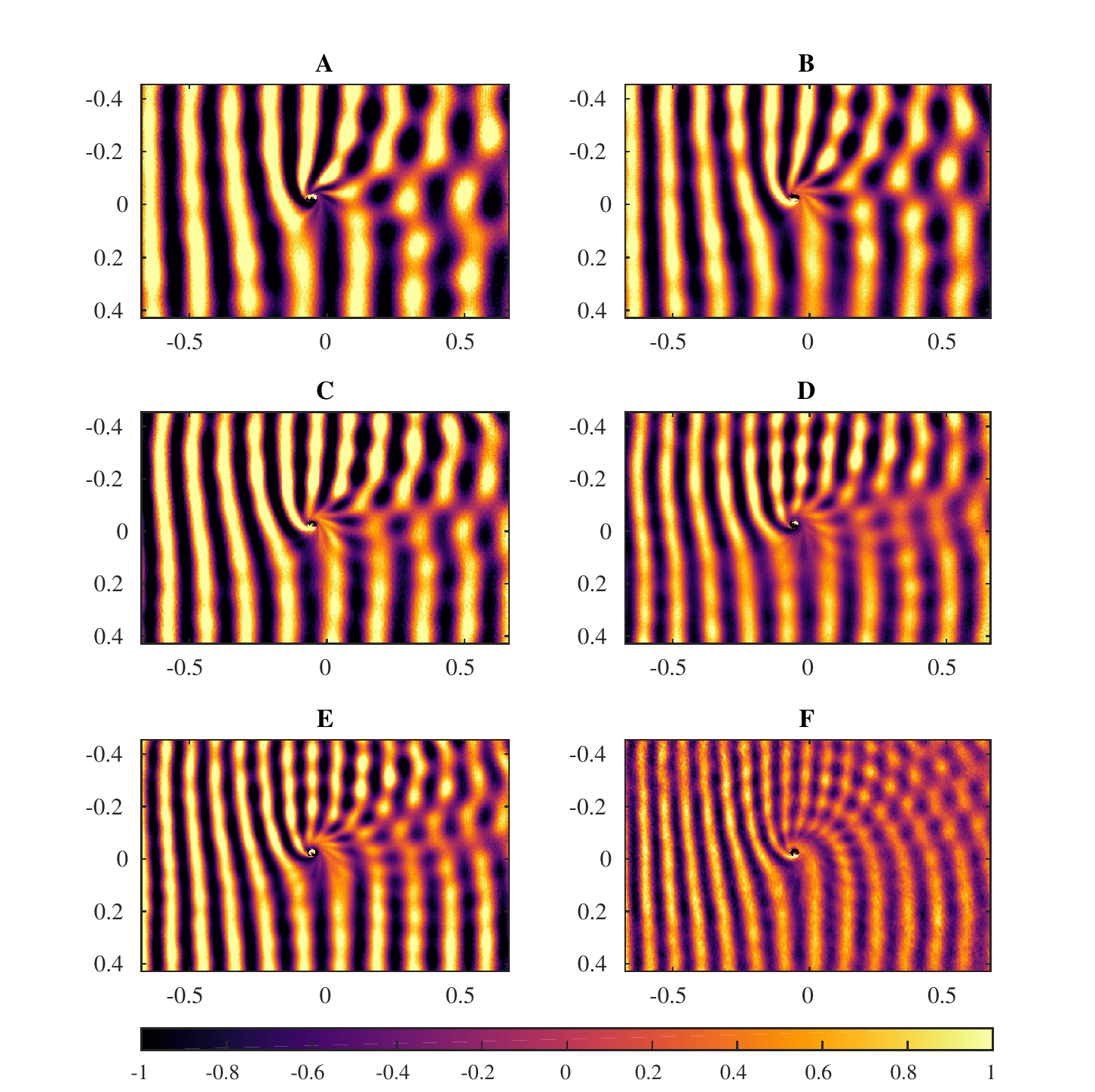} 
\end{center}
{\noindent \textbf{Figure 1 $|$} Wave characteristics of the surface perturbation $\xi$, filtered at a single frequency, for six different frequencies. The frequencies are $2.87 ~ \mathrm{Hz}$ \textbf{(A)}, $3.04 ~ \mathrm{Hz}$ \textbf{(B)}, $3.27 ~\mathrm{Hz}$ \textbf{(C)}, $3.45 ~\mathrm{Hz}$ \textbf{(D)}, $3.70 ~ \mathrm{Hz}$ \textbf{(E)}, and $4.11 ~\mathrm{Hz}$ \textbf{(F)}. The horizontal and vertical axis are in metres ($\mathrm{m}$), while the color scale is in millimetres ($\mathrm{mm}$). The patterns show the interfering sum of the incident wave with the scattered one. The waves are generated on the left side and propagate to the right across the vortex centred at the origin.} 

\begin{center} 
\includegraphics{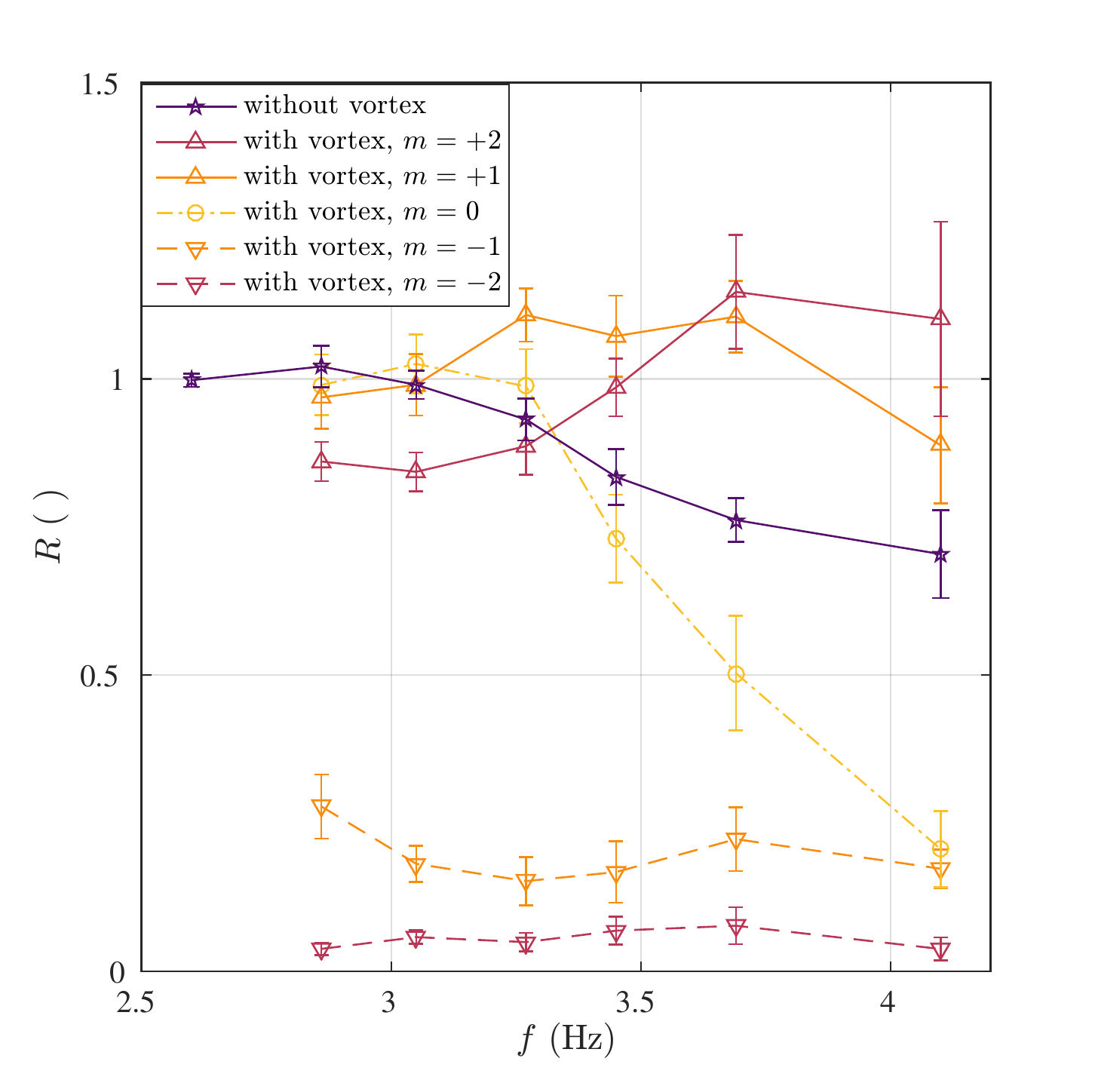}
\end{center}
{\noindent \textbf{Figure 2 $|$} Reflection coefficients for various frequencies and various $m$'s. For the vortex experiments the statistical average is taken over 6 repetitions, except for $f=3.70 ~\mathrm{Hz}$ where we have 15 repetitions. The purple line (star points) shows the reflection coefficients of a plane wave in standing water of the same height. We observe a significant damping for the frequencies above 3Hz (see Fig. 2). In future experiments, we hope to reduce this damping by working with purer water~\cite{Przadka2012}. Each point is a statistical average over the $m$'s over 5 repetitions. The errors bars include the standard deviation over these experiments, and the standard deviation over several centre choices (see Methods - Data analysis). 
} \label{Fig:spectra} 

\begin{center}
\includegraphics{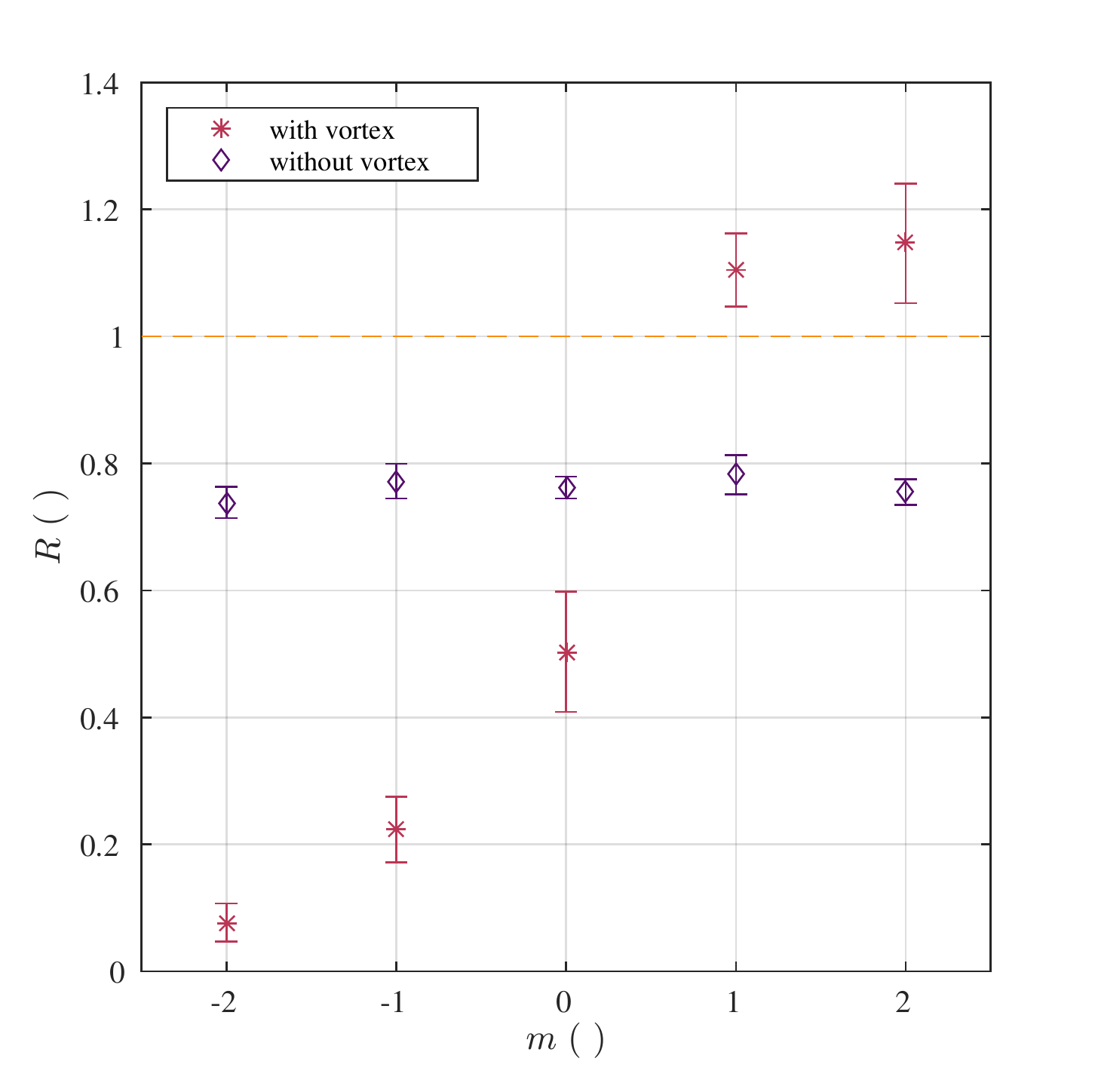} 
\end{center}
{\noindent \textbf{Figure 3 $|$} Reflection coefficients for different $m$'s, for the frequency $f = 3.70 ~ \mathrm{Hz}$ (stars). We have also shown the reflection coefficients for observed plane waves without a flow, at the same frequency and water height (diamonds). We see that the plane wave reflection coefficients are identical for all $m$'s, and all below 1 (within error bars). The statistic has been realized over 15 experiments. Error bars include the same contributions as in \textbf{Fig.~2}.\label{Fig:superradiance}} 

\begin{center}
\includegraphics{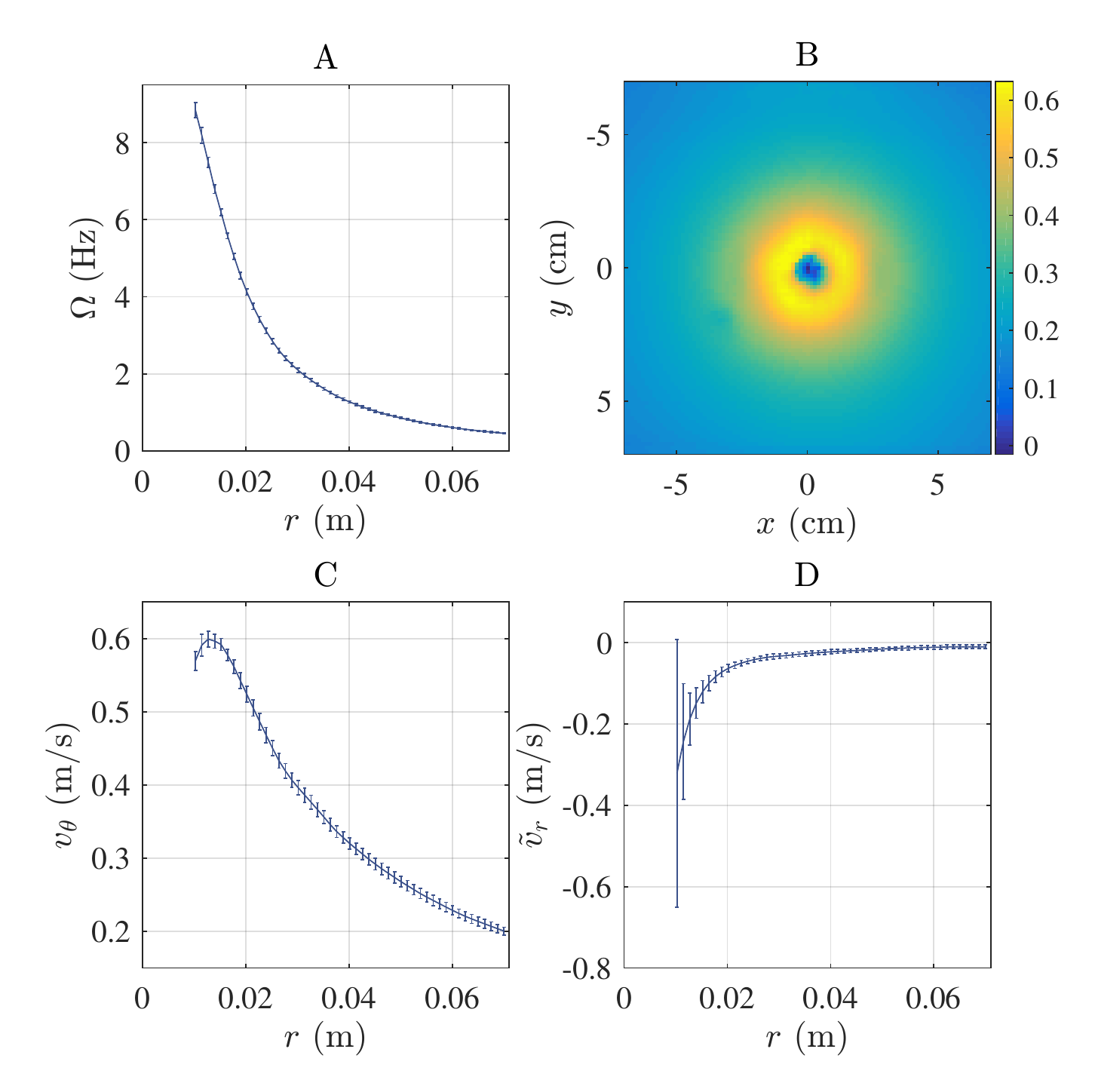} 
\end{center}
{\noindent \textbf{Figure 4 $|$} PIV measurements of the velocity field averaged of $10$ experiments. The error bars correspond to standard deviations across the measurements. \textbf{(A)} Angular frequency profile as a function of $r$. \textbf{(B)} Norm of the velocity field of the background flow. \textbf{(C)} $v_\theta$ profile as function of $r$.  \textbf{(D)} $\tilde v_r$ profile as function of $r$ (see Methods - PIV measurements).} \label{Fig:angularV} 

\begin{methods}
\subsection{Wave energy.} 
To verify that the observed amplification increases the energy of the wave, we compare the energy current of the inward wave with respect to the outward one. Since energy is transported by the group velocity $v_g$, the energy current is given by $J = g \, \om_{\rm fluid}^{-1} \, v_g |A|^2/f$ (up to the factor $1/f$, this is the wave action, an adiabatic invariant of waves~\cite{Buhler,Richartz:2012bd,Coutant:2016vsf}). If the background flow velocity is zero, then the ratio $J_{\rm out}/J_{\rm in}$ is simply $|A_{\rm out}/A_{\rm in}|^2$. However, in the presence of the vortex, we observe from our radial profiles $\varphi(r)$ (defined in equation \eqref{Fourier_t_th}) that the wave number of the inward and outward waves are not exactly opposite. The origin of this (small) difference is that the flow velocity is not completely negligible in the observation window. It generates a small Doppler shift that differs depending on whether the wave propagates against or with the flow. In this case, the ratio of the energy currents picks up a small correction with respect to the ratio of the amplitudes, namely, 
\be \label{Norm_eq}
 \frac{J_{\rm out}}{J_{\rm in}} = 
\left| \frac{\om_{\rm fluid}^{\rm in} v_g^{\rm out}}{\om_{\rm fluid}^{\rm out} v_g^{\rm in}}\right|  
\left|\frac{A_{\rm out}}{A_{\rm in}}\right|^2. 
\ee
To estimate this factor, we assume that the flow varies slowly in the observation window, such that $\om_{\rm fluid}$ obeys the usual dispersion relation of water waves, $\om_{\rm fluid}^2 = g k \tanh(h_0 k)$. (This amounts to a WKB approximation, and capillarity is neglected.) Under this assumption, the group velocity is the sum of the group velocity in the fluid frame, given by the dispersion relation, $v_g^{\rm fluid} = \p_k \sqrt{g k \tanh(h_0 k)}$, and the radial velocity of the flow $v_r$. Hence the group velocity needed for the energy ratio \eqref{Norm_eq} splits into two: $v_g = v_g^{\rm fluid} +v_r$. The first term is obtained only with the values of $k_{\rm in}$ and $k_{\rm out}$, extracted from the radial Fourier profiles. The second term requires the value of $v_r$, which we do not have to a sufficient accuracy. However, using the PIV data, we see that the contribution of this last term amounts to less than $1\%$ in all experiments.

\subsection{Data analysis.} 
We record the free surface of the water in a region of $1.33~ \mathrm{m} \times 0.98~ \mathrm{m}$ over the vortex during $13.2~ \mathrm{s}$. From the sensor we obtain $248$ reconstructions of the free surface. These reconstructions are triplets $X_{ij}$, $Y_{ij}$ and $Z_{ij}$ giving the coordinates of $640 \times 480$ points on the free surface. Because of the shape of the vortex, and noise, parts of the free surface cannot be seen by our sensor, resulting in black spots on the image. Isolated black spots are corrected by interpolating the value of the height using their neighbours. This procedure is not possible in the core of the vortex and we set these values to zero. 

To filter the signal in frequency, we first crop the signal in time so as to keep an integer number of cycles to reduce spectral leakage. We then select a single frequency corresponding to the excitation frequency $f_0$. After this filter, we are left with a 2-dimensional array of complex values, encoding the fluctuations of the water height $\xi(X_{ij},Y_{ij})$ at the frequency $f_0$. $\xi(X_{ij},Y_{ij})$ is defined on the grids $X_{ij}$, and $Y_{ij}$, whose points are not perfectly equidistant (this is due to the fact that the discretization is done by the sensor software in a coordinate system that is not perfectly parallel to the free surface).

To select specific azimuthal numbers, we convert the signal from cartesian to polar coordinates. For this we need to find the centre of symmetry of the background flow. We define our centre to be the centre of the shadow of the vortex, averaged over time (the fluctuations in time are smaller than a pixel). To verify that this choice does not affect the end result, we performed a statistical analysis on different centre choices around this value, and added the standard deviation to the error bars. Once the centre is chosen, we perform a discrete Fourier transform on the irregular grid $(X_{i j}, Y_{i j})$. We create an irregular polar grid $(r_{i j}, \theta_{i j})$ and we compute 
\begin{equation}
\varphi_{m}(r_{ij})= \frac{\sqrt{r_{ij}}}{2\pi} \sum_{j} \xi(r_{ij},\theta_{ij}) e^{-im\theta_{ij}} \Delta\theta_{ij},
\end{equation}
where $\Delta\theta_{ij} =(\Delta X_{ij} \Delta Y_{ij})/(r_{ij} \Delta r_{ij})$ is the line element along a circle of radius $r_{ij}$. 

To extract the inward and outward amplitudes $A_{\rm in}$ and $A_{\rm out}$, we compute the radial Fourier transform $\tilde \varphi_m(k) = \int \varphi_m(r) e^{-i k r} dr$ over the window $[r_{\rm min}, r_{\rm max}]$. Due to the size of the window compared to the wavelength of the waves, we can only capture a few oscillations in the radial direction, typically between 1 and 3. This results in broad peaks around the values $k_{\rm in}$ and $k_{\rm out}$ of the inward and outward components. We assume that these peaks contain only one wavelength (no superposition of nearby wavelengths), which is corroborated by the fact that we have filtered in time, and the dispersion relation imposes a single wavelength at a given frequency. To reduce spectral leakage, we use a Hamming window function on $[r_{\rm min}, r_{\rm max}]$, defined as 
\be
W(n)=0.54 - 0.46\cos \left(2\pi \frac{n}{N}\right), 
\ee
where $n$ is the pixel index running from $1$ to $N$. This window is optimized to reduce the secondary lobe, and allows us to better distinguish peaks with different amplitudes~\cite{Prabhu2013}. In \textbf{Extended Data Fig.~1}, 
we show the radial Fourier profiles for various $m$ for a typical experiment (left column), and the raw radial profiles and how they are approximated by Eq.~\eqref{in_out} (right column).

\subsection{PIV measurements.}

Close to the vortex core, the draining bathtub vortex is cylindrically symmetric to a good approximation. An appropriate choice of coordinates is, therefore, cylindrical coordinates $(r,\theta,z)$. The velocity field will be independent of the angle $\theta$ and can be expressed as
\begin{equation} \label{profile}
\textbf{v}(r,z) = v_r(r,z) \vec e_r + v_{\theta}(r,z) \vec e_\theta + v_z(r,z) \vec e_z.
\end{equation} 
We are specifically interested in the velocity field at the free surface $z = h(r)$. When the free surface is flat, $h$ is constant and the vertical velocity $v_z$ vanishes. When the surface is not flat, the $v_z$ component can be deduced from $v_r$ using the free surface profile $h(r)$ and the equation $v_z(r,h(r)) = (\partial_rh)v_r|_{z=h}$. To obtain an estimate of $v_z$, we use a simple model for the free surface shape~\cite{Lautrup2011},
\begin{equation}
h(r) = h_{0}\Big(1 - \frac{r_a^2}{r^2}\Big),
\end{equation}
where $h_{0}$ is the water height far from the vortex and $r_a$ is the radial position at which the free surface passes through the sink hole. This approximation captures the essential features of our experimental data. The components $v_r$ and $v_{\theta}$ are determined through the technique of Particle Imaging Velocimetry (PIV), implemented through the Matlab extension \textit{PIVlab} developed in Refs \cite{PIVthesis,PIVlab,PIVlab2}. The technique can be summarised as follows. 

The flow is seeded with flat paper particles of mean diameter $d = 2~ \mathrm{mm}$. The particles are buoyant which allows us to evaluate the velocity field exclusively at the free surface. The amount by which a particle deviates from the streamlines of the flow is given by the velocity lag $U_s = d^2(\rho-\rho_0)a/18\mu$ \cite{PIVthesis}, where $\rho$ is the density of a particle, $\rho_0$ is the density of water, $\mu$ is the dynamic viscosity of water and $a$ is the acceleration of a particle. For fluid accelerations in our system this is at most of the order $10^{-4}~ \mathrm{m/s}$, an order of magnitude below the smallest velocity in the flow. Thus we can safely neglect the effects of the velocity lag when considering the motions of the particles in the flow.

The surface is illuminated using two light panels positioned at opposite sides of the tank. The flow is imaged from above using a Phantom Miro Lab 340 high speed camera at a frame rate of $800~ \mathrm{fps}$ for an exposure time of $1200 ~\mathrm{\mu s}$. The raw images are analysed using \textit{PIVlab} by taking a small window in one image and looking for a window within the next image which maximizes the correlation between the two. By knowing the distance between these two windows and the time step between two images, it is possible to give each point on the image a velocity vector. This process is repeated for all subsequent images and the results are then averaged in time to give a mean velocity field. 

The resulting velocity field is decomposed onto an $(r,\theta)$-basis centred about the vortex origin to give the components $v_r$ and $v_{\theta}$. The centre is chosen so as to maximize the symmetry. In \textbf{Fig.~4B} 
we show the norm of the velocity field on the free surface. We see that our vortex flow is symmetric to a good approximation. To quantify the asymmetry of the flow, we estimate the coupling of waves with $m \neq m'$ through asymmetry. The change of the reflection coefficient due to this coupling is of the order of $|\tilde v^l/v_g|$, where $\tilde v^l$ is the angular Fourier component of azimuthal number $l=m-m'$. This ratio is smaller than $3\%$ in all experiments. To obtain the radial profiles of $v_r$ and $v_{\theta}$, we integrate them over the angle $\theta$. In \textbf{Figs.~4C} and  \textbf{4D} we show $v_{\theta}$ and the inward velocity tangent to the free surface $\tilde v_r = -\sqrt{v_{r}^2+v_{z}^2}$ as functions of $r$. The angular velocity of the flow is given by $\Omega(r) = v_{\theta}/r$ which is shown in \textbf{Fig.~4A}. 
From this plot it is clear that $\Omega$ reaches large enough values to be consistent with the detection of superradiance. 

\end{methods}



\begin{addendum}
 \item[Acknowledgements] We are indebted to the technical and administrative staff in the School of Physics \& Astronomy where our experimental setup is hosted. In particular, we want to thank Terry Wright and Tommy Napier for their support, hard work and sharing their technical knowledge and expertise with us to set up the experiment in Nottingham.
Furthermore we would like to thank Bill Unruh, Stefano Liberati, Joseph Niemela, Luis Lehner, Vitor Cardoso, Michael Berry, Vincent Pagneux, Daniele Faccio, Fedja Orucevic, J\"org Schmiedmayer and Thomas Fernholz for discussions regarding the experiment, and we wish to thank Michael Berry, Vitor Cardoso, Daniele Faccio, Luis Lehner, Stefano Liberati, and Bill Unruh for comments on the paper. Although all experiments have been conducted at The University of Nottingham, the initial stages of the experiment took place at ICTP\,/\,SISSA in Trieste (Italy) and would not have been possible without the support by Joseph Niemela, Stefano Liberati and Guido Martinelli. S.~W.~would like to thank Matt Penrice, Angus Prain, Miltcho Danailov, Ivan Cudin, Henry Tanner, Zack Fifer, Andreas Finke, and Dylan Russon for their contributions at different stages of the experiment. S.~W.~would also like to thank Thomas Sotiriou for the many discussions on all aspects of the project.
 
A.~C.~acknowledges funding received from the European Union's Horizon 2020 research and innovation programme under the Marie Sklodowska Curie grant agreement No 655524. M.~R.~acknowledges financial support from the S\~ao Paulo Research Foundation (FAPESP), Grants No.~2005/04219-0, No.~2010/20123-1, No.~2013/15748-0 and No.~2015/14077-0. M.~R.~and Ted T.~are also grateful to S.~W.~and the University of Nottingham for hospitality while this work was being completed. S.~W.~acknowledges financial support provided under the Royal Society University Research Fellow (UF120112), the Nottingham Advanced Research Fellow (A2RHS2), the Royal Society Project (RG130377) grants and the EPSRC Project Grant (EP/P00637X/1). The initial stages of the experiment were funded by S.~W.~'s Research Awards for Young Scientists (in 2011 and 2012) and by the Marie Curie Career Integration Grant (MULTI-QG-2011). 

 \item[Author Contributions] All authors contributed substantially to the work. 
 
 \item[Author Information] The authors declare that they have no competing financial interests. Correspondence and requests for materials
should be addressed to S.W.~(email: silke.weinfurtner@nottingham.ac.uk).
\end{addendum}


\begin{center}
\includegraphics{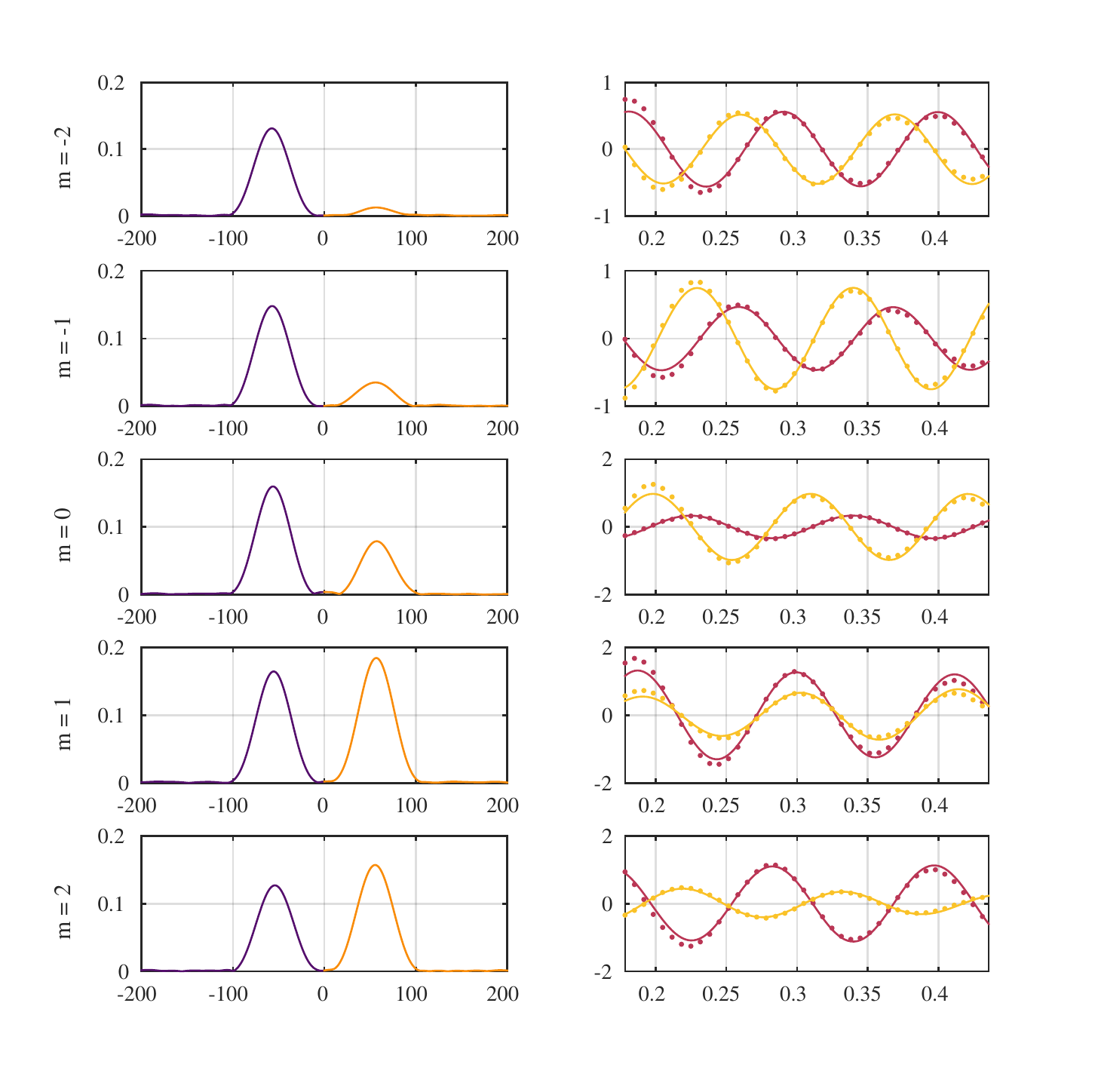} 
\end{center}
{\noindent \textbf{Extended Data Figure 1 $|$} Left side: Modulus of the Fourier profiles $|\tilde \varphi_m(k)|^2$ for various $m$. Right side: Radial profiles $\varphi_m(r)$ for various $m$ (maroon: real part, yellow: imaginary part). The vertical axis is in arbitrary units. The horizontal axes in inverse metres ($\mathrm{m^{-1}}$) on the left side, and metres ($\mathrm{m}$) on the right side. The dots are the experimental data (for clarity, only 1 out of 3 is represented), and the solid lines show the approximation of Eq.~\eqref{in_out} for the extracted values of $A_{\rm in}$ and $A_{\rm out}$.} \label{Fig:FourierAndReconstruction} 

\end{document}